\documentclass[prl,reprint,superscriptaddress]{revtex4-1}

\usepackage{graphicx}
\usepackage{amsmath}
\usepackage{amssymb}
\usepackage[titletoc]{appendix}
 
\usepackage{bm}
\usepackage{amsfonts}
\usepackage{epstopdf}
\usepackage{xcolor}
\usepackage{color}
\usepackage{wasysym}

\usepackage{booktabs}

\usepackage{verbatim}
\date{\today}

\begin{document}

\begin{abstract}

We introduce the first approach to volumetrically generate relativistically thermal plasma at gas-jet--accessible density. Using fully kinetic simulations and theory, we demonstrate that two stages of direct laser acceleration driven by two laser pulses in an applied magnetic field can heat a significant plasma volume to multi-MeV average energy. The highest-momentum feature is 2D-isotropic, persists after the interaction, and includes the majority of electrons, enabling experimental access to bulk-relativistic, high-energy-density plasma in an optically diagnosable regime for the first time.

\end{abstract}

\title{Underdense relativistically thermal plasma produced by magnetically assisted direct laser acceleration}

\author{K. Weichman}
\email[corresponding author, ]{kweic@lle.rochester.edu}
\affiliation{Laboratory for Laser Energetics, University of Rochester, Rochester, NY 14623, USA}
\affiliation{Department of Mechanical and Aerospace Engineering, University of California at San Diego, La Jolla, CA 92093, USA}

\author{J.P. Palastro}
\affiliation{Laboratory for Laser Energetics, University of Rochester, Rochester, NY 14623, USA}

\author{A.P.L. Robinson}
\affiliation{Central Laser Facility, STFC Rutherford-Appleton Laboratory, Didcot, OX11 0QX, UK}

\author{A.V. Arefiev}
\affiliation{Department of Mechanical and Aerospace Engineering, University of California at San Diego, La Jolla, CA 92093, USA}
\affiliation{Center for Energy Research, University of California at San Diego, La Jolla, CA 92037, USA}

\maketitle

The discovery of special relativity in 1905 transformed the fields of electromagnetism and charged particle kinetics that, some twenty years later, would coalesce into the field of plasma physics. Predictions have continually emphasized the importance of special relativity in plasmas where the majority of electrons are relativistic regardless of reference frame, but, even today, experimental verifications of these predictions remain relatively rare. The laboratory generation of these relativistically thermal plasmas is needed to address open questions in astrophysics regarding shock acceleration and the origin of cosmic rays~\cite{blandford1987shock}, fast radio bursts~\cite{bingham2003maser,metzger2019frb}, and $\gamma$-ray bursts~\cite{kumar2015grb}. 
Relativistically thermal plasmas also feature a substantially modified response to electromagnetic radiation relative to the nonrelativistic or nonthermal cases, which is of significant interest in basic plasma physics~\cite{bergman2001dispersion}, laboratory astrophysics~\cite{lontano2001pair,yang1994weibel}, and laser-plasma physics~\cite{stark2015polarizer,li2013hosing,zhao2014parametric,ross2010thomson}.

However, the production of relativistically thermal plasma in the laboratory with sufficient volume and duration for subsequent probing is challenging. 
Pulsed power and microwave sources, while capable of igniting thermal plasma over large volumes, are incapable of reaching relativistic electron temperatures. Laser pulses with relativistic intensity ($I_0 \gtrsim 10^{18}$~W/cm$^2$ for $\lambda_0 = 1\;\mu$m wavelength) are capable of imparting substantial energy to electrons, but are conventionally unable to create persistent, large-volume plasma where the majority of electrons are relativistic. 
Configurations involving opaque plasma ($n_\mathrm{e} > n_\mathrm{c}$, where $n_\mathrm{c}\approx 10^{21}\;\mathrm{cm}^{-3}$ is the critical density for $\lambda_0 = 1\;\mu$m)~\cite{purvis2013nanowire,weng2016rpa}, near-critical density plasma~\cite{li2008pschannel}, or acceleration by the plasma (wakefield) electric field~\cite{tajima1979,esarey2009lwfa} typically leave the majority of electrons cold either in momentum or configuration space.
In the underdense regime ($n_\mathrm{e}<n_\mathrm{c}$), laser pulses can volumetrically accelerate electrons to high energy~\cite{kruger1976dla,hartemann1995dla}, but the plasma does not remain hot after the laser pulse passes due to the reversibility of the acceleration process.
This reversibility is disrupted, however, by the addition of a uniform static magnetic field, enabling dramatic plasma heating.

In this Letter, we propose the first method to volumetrically generate relativistically-thermal, underdense plasma.
Our approach leverages two regimes of magnetically assisted direct laser acceleration (DLA). The energy retained following electron interaction with a short (femtosecond) laser pulse is used to catalyze subsequent heating by a long (picosecond) laser pulse, resulting in a 2D-isotropic momentum spectrum with relativistic average energy over a significant plasma volume. 
This result is robust to finite laser spot size in the magnetic-field direction, is predicted to occur over a wide range of laser and plasma conditions, and is accessible using current technology.
The heating process observed in 2D particle-in-cell (PIC) simulations is consistent with analytic modeling, which predicts that the combined short pulse, long pulse, and applied magnetic field produce multi-MeV plasma that persists for picoseconds following the interaction.

\begin{figure}
    \includegraphics[width=0.95\linewidth]{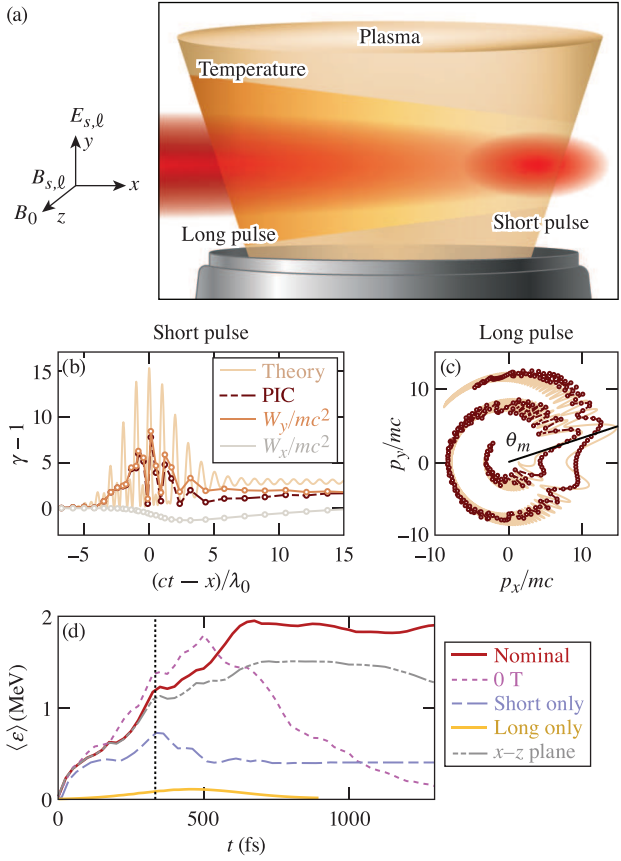}
    \caption{
    Generation of relativistic underdense plasma via magnetically assisted direct laser acceleration. (a)~Illustration of laser and magnetic-field configuration. [(b),(c)]~Example of the energy gain process for a representative electron interacting with (b)~the short pulse, and (c)~the long pulse. $W_y$ ($W_x$) is the work done by the transverse (longitudinal) electric field. (d)~Average energy of all electrons in $r < 25 \;\mu$m. Black dotted line: the time the peak of the short pulse leaves the plasma slab. The long pulse intensity has dropped to $a_\ell/\mathrm{e}$ at the right edge of the slab at the final time shown. The nominal case corresponds to both laser pulses and $B_{z0} = 500$~T, simulated in the $x$--$y$ plane. 
    }
    \label{fig:demo}
\end{figure}

Figures~\ref{fig:demo}(a)-(c) illustrate the two-stage configuration for generating relativistically thermal plasma. First, a $+x$-propagating, $y$-polarized relativistic short-pulse laser interacts with electrons in an underdense plasma with an embedded transverse magnetic field $B_0 \mathbf{\hat{z}}$, imparting net energy as electrons slip through the full pulse duration [Fig.~\ref{fig:demo}(b)]. Second, a longer laser pulse with the same propagation and polarization direction interacts with these pre-heated electrons, delivering half-laser-cycle energy kicks that promote the electron to higher-energy cyclotron orbits [Fig.~\ref{fig:demo}(c)]. The result of these two interactions is a large volume of underdense plasma heated to relativistic energy. 

\begin{table}
\centering
\begin{tabular}{ |l|l| }
  \hline
  \multicolumn{2}{|l|}{\textbf{Laser parameters} }\\
  \hline
  Laser polarization & $y$ \\
  Propagation direction & $+x$ \\
  Wavelength & $\lambda_0=1$ $\mu$m \\
  Spot size (Gaussian, FWHM in $|E|$) & $100$ $\mu$m\\
  \hline
  \multicolumn{2}{|l|}{\textbf{Short pulse} } \\
  Duration (Gaussian, FWHM in $|E|$) & $\tau_s = 20$ fs\\
  Peak amplitude & $a_s = 5$ \\
  \hline
  \multicolumn{2}{|l|}{\textbf{Long pulse} } \\
  Duration (Gaussian, FWHM in $|E|$) & $\tau_\ell = 0.8$ ps\\
  Peak amplitude & $a_\ell =1$ \\ 
  Delay relative to short pulse & $0.6 \tau_\ell$ \\
  \hline \hline
  \multicolumn{2}{|l|}{\textbf{Other parameters} }\\
  \hline
  Applied magnetic field ($\mathbf{B}=B_0 \mathbf{\hat{z}}$) & $B_0 = 500$ T \\  
  Plasma length & $L = 100$ $\mu$m\\
  Electron density & $n_\mathrm{e} = 10^{-3}\;n_\mathrm{c}$ \\  
  Location of plasma surface & $x=0$ \\
  Time when peak of short pulse is at $x=0$ & $t = 0$ \\
  Simulation plane & $x$--$y$ \\
  Spatial resolution & 30 cells/$\lambda_0$ \\
  Macroparticles per cell, electron/proton &  10/5 \\ 
  \hline
   \end{tabular}
  \caption{
  Nominal 2D simulation parameters. The delay between pulses corresponds to long-pulse amplitude $a_\ell/\mathrm{e}$ at the peak of the short pulse. Simulations were conducted with high-order cubic B-spline particle shape (which produces robust energy conservation) using the open source particle-in-cell code \textit{EPOCH}~\cite{arber2015epoch}.
  }
  \label{table:parameters}
\end{table}

We demonstrate the realization of this novel hot plasma regime in 2D particle-in-cell simulations. 
For illustrative purposes, we construct a nominal case using a gas-jet--relevant hydrogen plasma (density $10^{-3}n_\mathrm{c}\approx 10^{18}\;\mathrm{cm}^{-3}$) and weakly focused laser pulses ($f_\#\sim 90$). The short (subscript $s$) and long (subscript $\ell$) laser pulses are weakly to moderately relativistic with a peak normalized electric-field amplitude ($a_0 = |e|E_0/mc\omega_0$, where $\omega_0$ is the laser frequency) of $a_s = 5$ and $a_\ell = 1$. 
Other parameters are given in Table~\ref{table:parameters}. 

\begin{figure}
    \includegraphics[width=0.95\linewidth]{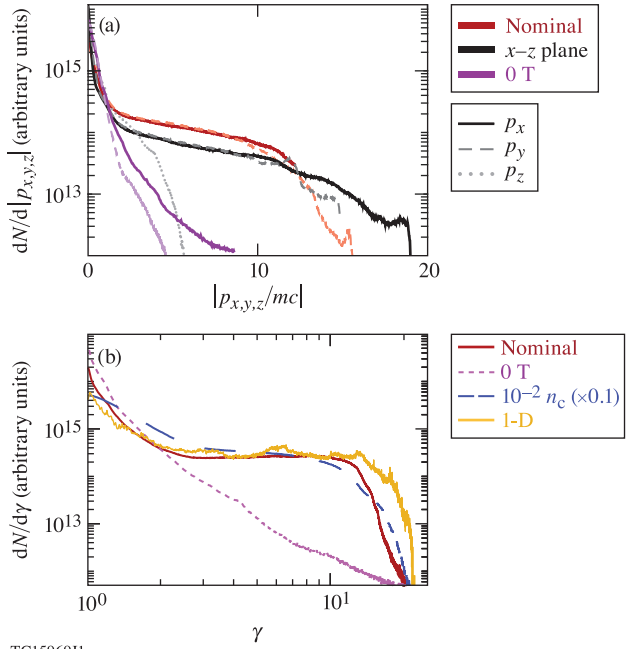}
    \caption{
    Characteristic spectra produced by magnetically assisted direct laser acceleration for (a)~momentum in each direction and (b)~$\gamma$. Spectra correspond to $r<25\;\mu$m and the final time in Fig.~\ref{fig:demo}(d).
    }
    \label{fig:spectra}
\end{figure}

The interaction of the two laser pulses with the target creates multi-MeV average electron energy over a large volume (e.g., $r<w/2 = 25\;\mu$m, where $w$ is the HWHM laser spot size), which persists for picoseconds following the interaction [Fig.~\ref{fig:demo}(d)]. 
Despite the strong heating, the density remains close to its starting value ($n_\mathrm{e} \approx 0.9 \pm 0.3 \times 10^{-3} n_\mathrm{c}$).
The corresponding momentum spectrum is 2D-isotropic [in $p_x$ and $p_y$, Fig.~\ref{fig:spectra}(a)], with a flat energy spectrum [Fig.~\ref{fig:spectra}(b)]. While the plasma can be heated somewhat by the short laser pulse and magnetic field alone, significant relativistic heating requires all three elements of the short laser pulse, long laser pulse, and applied magnetic field [c.f., cases in Fig.~\ref{fig:demo}(d)]. 
Unlike conventional laser-based heating methods, the flat, relativistic feature ($\gamma \gtrsim 2$) includes more than half of the electron population, i.e., the plasma is relativistically thermal.

These observations are explainable as volumetric heating by magnetically assisted direct laser acceleration in the two distinct regimes covered by the short pulse and the long pulse.
While conventional direct laser acceleration in a plane wave (i.e., with $B_0 = 0$) is reversible, the addition of a weak magnetic field transverse to the laser propagation direction slowly rotates the forward electron momentum into transverse momentum and introduces multi-cycle and half-cycle opportunities for irreversible acceleration. 
During electron interaction with the short pulse, momentum rotation by the applied magnetic field changes the dephasing rate $R=\gamma - p_x/mv_\phi$ (where $v_\phi$ is the phase velocity) over many cycles, allowing the electron to retain energy after the laser has passed~\cite{robinson2020dephasing}. 
Substantial momentum rotation also occurs during a single half laser cycle when the interaction is sufficiently long~\cite{arefiev2020dla}, and can impart higher net energy than the multi-cycle mechanism.
However, entering the necessary regime for half-cycle magnetically assisted DLA by the long pulse requires initial heating, e.g., preheating by the short pulse. 

To demonstrate the need for the short pulse to catalyze heating by the long pulse, we write the electron equations of motion 
\begin{equation} \label{eqn:basic}
\dfrac{\mathrm{d} \mathbf{p}}{\mathrm{d}t} = -|e| \mathbf{E} - \dfrac{|e|\mathbf{p}}{\gamma m c} \times \mathbf{B}
\end{equation}
in terms of the evolution of the angle the electron momentum makes with the forward direction, $p_x = |p| \cos \theta$, and the electron energy,
\begin{align}
        & \dfrac{\mathrm{d}\gamma}{\mathrm{d}s} = \dfrac{|p_y|}{p_\perp} \dfrac{ \beta \sin \theta}{1- \left(\beta/\beta_\phi\right) \cos \theta} \dfrac{\mathrm{d}a}{\mathrm{d}s}, \label{eqn:gam_s} \\
        & \dfrac{\mathrm{d} \theta}{\mathrm{d} s} = \dfrac{|p_y|}{p_\perp} \dfrac{ \dfrac{\omega_{\mathrm{c}0}}{\omega_0} + \dfrac{1}{\beta}\left[\cos \theta - \beta/\beta_\phi  \right] \dfrac{\mathrm{d}a}{\mathrm{d}s}}{\gamma \left[1- \left(\beta/\beta_\phi\right)  \cos \theta\right]}, \label{eqn:angle}         
\end{align}
where $s = \omega_0 (t - x/v_\phi)$ is the laser phase variable, $a$ is the normalized vector potential, $\omega_{\mathrm{c}0} = |e|B_0/m$ is the (nonrelativistic) cyclotron frequency, $p_\perp = \sqrt{p_y^2 + p_{z0}^2}$ ($p_z$ is constant), $\beta_\phi = v_\phi/c$, and $\beta = |p|/\gamma mc$.

Without the applied magnetic field, there is an angle $\theta_*$ given by 
$\cos \theta_* = \beta/\beta_\phi$
which results in $\mathrm{d}\theta/\mathrm{d}s=0$. While $\theta_*$ also maximizes the rate of energy gain [Eq.~(\ref{eqn:gam_s})], the energy imparted during the accelerating half-cycle is exactly removed during the subsequent, decelerating half-cycle, resulting in no net acceleration.

The addition of an applied magnetic field modifies the angle for which $\mathrm{d}\theta/\mathrm{d}s = 0$ to~\cite{arefiev2020dla}
\begin{equation} \label{eqn:theta_a}
    \theta \approx \sqrt{2\left(1 - \dfrac{\beta}{\beta_\phi}\right) + 2\beta \dfrac{\omega_{\mathrm{c}0}}{a_\ell \omega_0}} = \sqrt{\theta_*^2 + \theta_m^2},
\end{equation} 
in the small angle limit, with $\mathrm{d}a/\mathrm{d}s \sim a_\ell$ and $\theta_m \equiv \sqrt{2\beta \omega_{\mathrm{c}0}/a_\ell\omega_0}$. 
When $\theta_m \gtrsim \theta_*$,
Eqs.~(\ref{eqn:gam_s}) and~(\ref{eqn:angle}) predict half-cycle acceleration with the constant angle $\theta_m$, in good agreement with both observed electron trajectories and those calculated from the solution to Eq.~(\ref{eqn:basic}) in the vacuum plane wave limit [Fig.~\ref{fig:demo}(c)]. 

However, $\theta_m \gtrsim \theta_*$ requires electrons be preheated prior to the interaction. 
The initial $\gamma_0$ needed to catalyze half-cycle acceleration is approximately
\begin{equation} \label{eqn:emin_b}
    \gamma_0 \gtrsim f \sqrt{\dfrac{a_\ell}{2}\dfrac{\omega_0}{\omega_{\mathrm{c}0}}}, 
\end{equation}
where $f$ is given by $f = \exp{\left(-2a_\ell f\right)}$, and accounts for acceleration near $\theta=0$ (see Supplemental Material). For $a_\ell = 1$, $\gamma_0 \gtrsim 0.3 \sqrt{\omega_0/\omega_{\mathrm{c}0}}$.

\begin{figure}
    \includegraphics[width=0.95\linewidth]{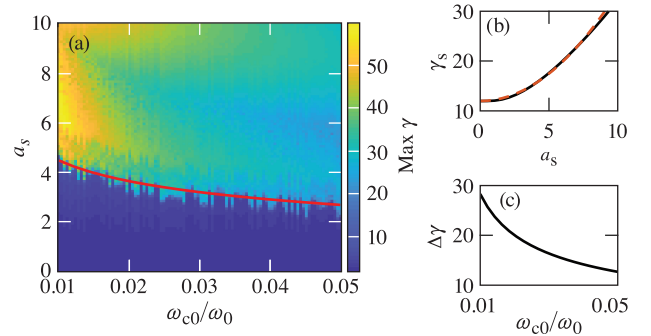}
    \caption{
    Threshold for energy gain based on model calculations. (a)~Calculated maximum $\gamma$ with $a_\ell = 1$. Red curve: $\gamma_s = \gamma_0$, where $\gamma_s$ is the energy retained following interaction with the short pulse, and $\gamma_0$ is given by Eq.~(\ref{eqn:emin_b}). (b)~Black: calculated value of $\gamma_s$ for the optimum short pulse duration ($\gamma_s$ is independent of $\omega_{\mathrm{c}0}/\omega_0$). Red: fit $\gamma_s \approx 1 + 0.11 a_s^2$. (c)~$\Delta \gamma$ for a single energy kick [Eq.~(\ref{eqn:gam_c})]. In (a), the short-pulse duration was scaled with $(\omega_{\mathrm{c}0}/\omega_0)^{-1}$ to maintain the optimal $\tau_s$ and the long-pulse duration was scaled with $(\omega_{\mathrm{c}0}/\omega_0)^{-3/2}$ to keep $\tau_\ell/\tau_L$ constant, where $\tau_L$ is the (relativistic) cyclotron period associated with $\Delta \gamma$.
    }
    \label{fig:theory}
\end{figure}

In the two-laser configuration, preheating is provided by the short pulse through many-cycle magnetically assisted DLA~\cite{robinson2020dephasing} [illustrated for a single electron in Fig.~\ref{fig:demo}(b)]. 
Eq.~\ref{eqn:emin_b} introduces a threshold in the short pulse intensity needed to catalyze heating by the long pulse, which is illustrated in the vacuum plane wave limit [solution to Eq.~(\ref{eqn:basic}), scanning over the initial laser phase] in Figs.~\ref{fig:theory}(a) and~\ref{fig:theory}(b).
The $a_s$ used in simulations is significantly above the observed threshold, which accommodates the detrimental effect of wakefield formation on acceleration by the short pulse [via increased rate of phase slip, see work done by longitudinal and transverse electric fields in Fig.~\ref{fig:demo}(b)]. 
Substantial plasma heating only occurs when the short pulse is included [nominal vs picosecond-only cases in Fig.~\ref{fig:demo}(d)].

The maximum half-cycle energy kick the long pulse can provide is~\cite{arefiev2020dla}
\begin{equation} \label{eqn:gam_c}
    \Delta \gamma \sim 2^{3/2} \, a_\ell^{3/2} \sqrt{\dfrac{\omega_0}{\omega_{\mathrm{c}0}}},
\end{equation}
shown in Fig.~\ref{fig:theory}(c), where this estimate was obtained by integrating Eq.~(\ref{eqn:gam_s}) over half a laser cycle in the small angle limit with $\theta \approx \theta_m$, $\beta \approx 1$, and $p_{z0}=0$.
The energy imparted by the accelerating half-cycle is at least partly retained during subsequent laser cycles [the rate of energy transfer in Eq.~(\ref{eqn:gam_s}) decreases with increasing $\theta$]. Eventually cyclotron rotation returns the electron to favorable conditions for energy exchange where it can get another large kick from the laser, either promoting the electron to a yet-higher energy orbit [as can be seen in Fig.~\ref{fig:demo}(c)] or resulting in energy loss. 
Although electrons typically undergo more than one half-cycle energy kick in the nominal simulation, the cutoff energy (where $\mathrm{d}N/\mathrm{d}\gamma$ drops rapidly) remains comparable to $\Delta \gamma$ [Fig.~\ref{fig:spectra}(b)].

While 2D simulation demonstrates the feasibility of producing multi-MeV relativistically thermal plasma under experimentally relevant conditions, the parameters chosen for these simulations (to maintain tolerable computational cost) produce sub-optimal plasma heating. 
One-dimensional simulation yields a nearly identical hot-electron spectrum as the 2D case [$\gamma \gtrsim 2$ in Fig.~\ref{fig:spectra}(b); i.e., the heating mechanism is fundamentally 1D-like] but overpredicts the average electron energy by $\sim 2\times$. This difference is linked to a smaller population of cold return current electrons in the 1D case (in 2D, return current electrons primarily originate at radii outside the central spot). Nevertheless, 1D simulations indicate that even hotter plasma can be generated by increasing the plasma size and long-pulse duration, and that heating can be facilitated by weaker applied magnetic fields.

\begin{figure}
    \centering
    \includegraphics[width=0.95\linewidth]{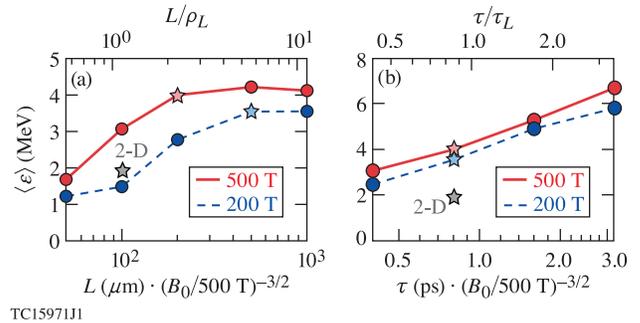}
    \caption{Strategies for improving average electron energy in 1D PIC simulations. (a)~Scan over plasma size near $L/\rho_L \sim 1$ with fixed duration. (b)~Scan over long-pulse duration near $\tau_\ell/\tau_L \sim 1$ with fixed plasma size.  $\rho_L$ and $\tau_L$ are the Larmor radius and cyclotron period associated with $\Delta \gamma$. The starred points are shared between (a)~and~(b). The peak of the short pulse is kept coincident with $a_\ell/\mathrm{e}$ on the rising edge of the long pulse. $\tau_s = 50$~fs for the 200-T cases.
    }
    \label{fig:1D}
\end{figure}

First, increasing the plasma size increases the average electron energy by reducing electron interaction with the plasma--vacuum boundary. 
The plasma size in 2D simulations $L=100\;\mu\mathrm{m}$ is comparable to the Larmor radius of hot electrons $\rho_L \sim c\Delta \gamma/\omega_{\mathrm{c}0}$, meaning a significant portion of electrons encounter the decelerating sheath electric field during their acceleration, which is detrimental to the half-cycle energy gain process. 
Increasing the plasma size results in saturation of the average electron energy at a few times the Larmor radius, as shown in Fig.~\ref{fig:1D}(a).

Second, increasing the pulse duration increases the number of half-cycle energy kicks electrons receive, which increases the cutoff energy of the flat spectral feature.
The long-pulse duration used in 2D simulations is comparable to the cyclotron period of hot electrons ($\tau_\ell \sim \tau_L$, where $\tau_L \approx 2 \pi \rho_L/c$), allowing most electrons to receive one or two significant energy kicks. As the pulse duration is increased, additional kicks are allowed, increasing both the cutoff and the average electron energy [Fig.~\ref{fig:1D}(b)]. 

Lastly, relativistically thermal plasma can still be produced when the magnetic field is reduced to 200~T, which will enable the use of both pulsed power~\cite{portugall1999field_gen,ivanov2018zebra,fiksel2015mifeds,fiksel2018mifeds,mifedscomm} and laser-driven~\cite{gao2016coil,goyon2017coil,fujioka2013coil,santos2018coil} devices. Simulations conducted with a 200-T magnetic field and scaled pulse durations ($\omega_{\mathrm{c}0} \tau_s$ and $\tau_\ell/\tau_L$ kept constant on decreasing $B_0$) exhibit similar saturation behavior with increasing $L$ [Fig.~\ref{fig:1D}(a)] and similar scaling with pulse duration in the saturated regime [Fig.~\ref{fig:1D}(b)] as the 500-T case. 
In agreement with Eq.~(\ref{eqn:gam_c}), the magnetically assisted DLA spectral feature extends to higher energy in the 200-T case. However, the average electron energy is somewhat lower due to fewer electrons meeting the pre-acceleration requirement for the same $a_s$.
Higher $\gamma_0$ is required to kickstart the half-cycle acceleration process and stronger wakefield generation (from decreasing $B_0$ and increasing the pulse duration) lowers the average energy retained following electron interaction with the short pulse.

The two-pulse magnetically assisted DLA process is robust to additional experimental considerations, including increasing the plasma density by an order of magnitude [2D $10^{-2}\;n_\mathrm{c}$ case in Fig.~\ref{fig:spectra}(b)] and finite laser spot size in the magnetic field ($z$) direction. 
As expected from Eqs.~(\ref{eqn:gam_s}) and~(\ref{eqn:angle}), 2D simulation in the $x$--$z$ plane (which introduces electron motion along magnetic field lines) features the same acceleration process as the nominal case and also produces a 2D-isotropic momentum spectrum [Fig.~\ref{fig:spectra}(a)], albeit with somewhat reduced average energy [e.g., Fig.~\ref{fig:demo}(d)].  
The finite spot size in the magnetic-field direction additionally sets a picosecond time scale for dissipation of the hot plasma (based on $100$-$\mu$m-scale spot size and the observed $\langle p_z \rangle/\langle \gamma \rangle mc \approx 0.16$).

As a final note, the two-pulse scheme is advantageous relative to plasma heating by the short pulse alone. 
$\Delta \gamma$ is higher than the energy retained following interaction with the short pulse ($\gamma_s$), even if the combined pulse energy is given entirely to the short pulse [i.e., $a_s=8$, compare Figs.~\ref{fig:theory}(b) and~\ref{fig:theory}(c)].
In both 1D and 2D simulations, both the cutoff and average electron energy are substantially higher with the two-pulse scheme (c.f., average energy of 1.9~MeV versus 1.1~MeV for the nominal 2D case), and the characteristic flat energy feature [shown in Fig.~\ref{fig:spectra}(b)] is only observed when both pulses are present.

Our 1D and 2D simulations demonstrate the generation of underdense, relativisitically thermal plasma can be realized with currently available laser and magnetic-field--generation capabilities. 
With a 200-T magnetic field, we anticipate multi-MeV average electron energy under gas-jet relevant conditions ($n_\mathrm{e} \sim 10^{18}\;\mathrm{cm}^{-3}$, few-millimeter plasma size) using kilojoule-class laser pulses with a few hundred $\mu$m spot size and 50~fs/multipicosecond duration. 
Our approach is thereby anticipated to offer the first practical access to the relativistically thermal plasma regime, enabling experimental verification of long-standing, foundational predictions in basic plasma physics, laboratory astrophysics, and laser-plasma physics.

\section*{Acknowledgements}

We thank R. Bingham (STFC Rutherford-Appleton Laboratory) for useful discussions.
This material is based upon work supported by the Department of Energy National Nuclear Security Administration under Award Number DE-NA0003856, the University of Rochester, and the New York State Energy Research and Development Authority, and the DOE Office of Science under Grant No. DESC0018312. A.V.A. was supported by NSF Grant No. 1903098. The support of DOE does not constitute an endorsement by DOE of the views expressed in this paper.
Particle-in-cell simulations were performed using EPOCH \cite{arber2015epoch},  developed under UK EPSRC Grant Nos. EP/G054940, EP/G055165, and EP/G056803.
This work used HPC resources of the National Energy Research Scientific Computing Center (NERSC), a U.S. Department of Energy Office of Science User Facility operated under Contract No. DE-AC02-05CH11231, and the Extreme Science and Engineering Discovery Environment (XSEDE) \cite{towns2014xsede}, which is supported by National Science Foundation grant number ACI-1548562, under allocation TG-PHY190034 on the Texas Advanced Computing Center (TACC) at The University of Texas at Austin.
	
This report was prepared as an account of work sponsored by an agency of the U.S. Government. Neither the U.S. Government nor any agency thereof, nor any of their employees, makes any warranty, express or implied, or assumes any legal liability or responsibility for the accuracy, completeness, or usefulness of any information, apparatus, product, or process disclosed, or represents that its use would not infringe privately owned rights. Reference herein to any specific commercial product, process, or service by trade name, trademark, manufacturer, or otherwise does not necessarily constitute or imply its endorsement, recommendation, or favoring by the U.S. Government or any agency thereof. The views and opinions of authors expressed herein do not necessarily state or reflect those of the U.S. Government or any agency thereof.



\end{document}


\begin{abstract}
\end{abstract}

\title{Supplemental Material: Underdense relativistically thermal plasma produced by magnetically assisted direct laser acceleration}

\author{K. Weichman}
\affiliation{Laboratory for Laser Energetics, University of Rochester, Rochester, NY 14623}
\affiliation{Department of Mechanical and Aerospace Engineering, University of California at San Diego, La Jolla, CA 92093, USA}
\email[corresponding author, ]{kweic@lle.rochester.edu}
\author{J.P. Palastro}
\affiliation{Laboratory for Laser Energetics, University of Rochester, Rochester, NY 14623}
\author{A.P.L. Robinson}
\affiliation{Central Laser Facility, STFC Rutherford-Appleton Laboratory, Didcot, OX11 0QX, UK}
\author{A.V. Arefiev}
\affiliation{Department of Mechanical and Aerospace Engineering, University of California at San Diego, La Jolla, CA 92093, USA}
\affiliation{Center for Energy Research, University of California at San Diego, La Jolla, CA 92037, USA}

\maketitle

\section{Estimated energy required to kickstart half-cycle acceleration}

Observing the effect of the magnetic field on direct laser acceleration requires $\theta_m \gtrsim \theta_*$ at the start of the accelerating half-cycle. 
We approximate the required energy $\gamma_i$ to satisfy $\theta_m \gtrsim \theta_*$ with the simplifications that 
$1-\beta \gg \beta_\phi -1$ (which allows $\beta_\phi \to 1$,
valid for underdense, moderately relativistic plasma) and $\theta_m \approx \sqrt{2 \omega_{\mathrm{c}0}/a_\ell\omega}$ (i.e., $\beta \to 1$ in $\theta_m$). The required energy prior to the accelerating half-cycle is therefore 
\begin{equation}\tag{S.1}
    \gamma_i \gtrsim \sqrt{a_\ell \omega_0/2\omega_{\mathrm{c}0}}, 
\end{equation}
where we denote this energy as $\gamma_i$ to distinguish it from $\gamma_0$, which is the externally provided energy required to catalyze the heating. $\gamma_0$ is less than $\gamma_i$ because the electron can be accelerated not only near $\theta \approx \theta_m$, but also during the transit through $\theta = 0$, as occurs for the electron in Fig.~(1)c. Net acceleration about $\theta = 0$ is allowed when the sign of $\theta$ changes at a different time than the sign of $\mathrm{d}a/\mathrm{d}s$, such that the energy change [Eq.~(2)] during the accelerating and decelerating parts of the laser cycle do not exactly cancel.

To provide an estimate for $\gamma_0$, we consider the upper limit on the energy gain ($\gamma_i-\gamma_0$) 
that can occur during the crossing through $\theta = 0$.
Equation~(2) can also be written as 
\begin{equation}\tag{S.2}
R \dfrac{\mathrm{d}\gamma}{\mathrm{d} s} = \dfrac{|p_y|}{p_\perp} \left(\gamma \sin \theta \dfrac{\mathrm{d} a}{\mathrm{d} s} \right).   
\end{equation}
We take the upper limit in all terms on the right hand side ($\sin \theta \lesssim \theta_m$, $\beta \lesssim 1$, $|p_y|/p_\perp \lesssim 1$, $\mathrm{d}a/\mathrm{d}s \lesssim a_\ell$) and assume the minimum $R_\mathrm{min} \approx \gamma_0 - \sqrt{\gamma_0^2-1} \approx 1/2\gamma_0$. With these substitutions, we have $\mathrm{d}\gamma /\mathrm{d} s \lesssim \gamma a_\ell\theta_m/R_\mathrm{min}$. Integrating this expression with $\theta_m = \theta_* = 1/\gamma_i$ gives $\log(\gamma_i/\gamma_0) \lesssim 2 a_\ell (\gamma_0/\gamma_i)$. The required initial $\gamma_0$ for energy gain is therefore approximately
\begin{equation}\tag{S.3}
    \gamma_0 \gtrsim f \sqrt{\dfrac{a_\ell}{2}\dfrac{\omega_0}{\omega_{\mathrm{c}0}}}, \label{eqn:emin_b}
\end{equation}
where $f$ is the solution to $f = \exp{\left(-2a_\ell f\right)}$.